%% file: main.tex
\documentclass[sigconf,screen,nonacm]{acmart}
\AtBeginDocument{%
  }

\setcopyright{acmlicensed}
\copyrightyear{2026}
\acmYear{2026}
\acmDOI{XXXXXXX.XXXXXXX}
\acmConference[FutureHCI '26]{The Future of HCI Workshop}{August 17--18, 2026}{Blacksburg, VA, USA}
\acmISBN{978-1-4503-XXXX-X/2018/06}

\input{misc/usepackage}

\begin{document}

\title{Generating Synthetic Behavioral Populations from XR Motion}

\author{Xiaozheng Wang}
\orcid{0009-0004-6886-294X}
\affiliation{%
  \institution{Virginia Tech}
  \city{Blacksburg}
  \state{Virginia}
  \country{USA}
}
\email{xzwang@vt.edu}

\author{Ryan P. McMahan}
\orcid{0000-0001-9357-9696}
\affiliation{%
  \institution{Virginia Tech}
  \city{Blacksburg}
  \state{Virginia}
  \country{USA}}
\email{rpm@vt.edu}

\renewcommand{\shortauthors}{Wang et al.}

\begin{abstract}
Large-scale behavioral datasets are becoming increasingly important for machine learning, personalization, and behavioral modeling in extended reality (XR). However, collecting XR motion data from hundreds or thousands of participants remains expensive, time-consuming, and difficult to reproduce across research groups. As a result, many XR studies continue to rely on relatively small datasets that limit the scale and diversity of behavioral evaluation. To address this limitation, we investigate synthetic behavioral populations as a complementary approach to traditional XR data collection. We present an interpolation-based motion synthesis pipeline that combines dynamic time warping (DTW) with trajectory interpolation to generate synthetic behavioral trajectories from existing XR datasets while preserving task structure and incorporating motion characteristics from contributing participants. Using the publicly available FAST VR assembly dataset, we generated and openly released 100 synthetic behavioral trajectories. We evaluated the synthesized trajectories through motion-based user identification. Hybrid datasets containing both real and synthesized trajectories achieved performance comparable to similarly sized real-only datasets while maintaining low confusion between synthesized trajectories and their contributing participants. Rather than serving as conventional data augmentation, the proposed approach generates distinguishable behavioral trajectories that expand XR behavioral populations for larger-scale behavioral modeling and machine learning evaluation. Our findings demonstrate that synthetic behavioral populations provide a promising approach to expanding XR behavioral datasets and supporting future data-driven immersive systems.

\end{abstract}

\ccsdesc[500]{Human-centered computing~Virtual reality}
\ccsdesc[500]{Human-centered computing~User models}
\ccsdesc[500]{Computing methodologies~Machine learning}


\keywords{VR, Machine Learning, Identification}


\maketitle
\input{Content/2-Introduction}
\input{Content/3-Related_Work}

\input{Content/4-Identity-Level_Data_Synthesis}

\input{Content/5-Machine_Learning_Experiment}

\input{Content/6-Study}
\input{Content/7-Discussion}
\input{Content/8-Limitation_and_Future_Work}
\input{Content/9-Conclusion}

\balance
\bibliographystyle{ACM-Reference-Format}
\bibliography{ref}

\end{document}

%% file: misc/usepackage.tex
\graphicspath{{figures/}{pictures/}{images/}{./}} 

\usepackage{booktabs}                  

\usepackage{graphicx}
\usepackage{makecell}
\usepackage{array}

\usepackage{amsmath}
\usepackage{xcolor}
\usepackage{enumitem}

\usepackage{pifont}
\usepackage{tabularx} 

\usepackage{xcolor}

\usepackage{ragged2e}

\newcolumntype{Y}{>{\RaggedRight\arraybackslash}X}
\newcolumntype{Z}{>{\centering\arraybackslash}X}

\usepackage{threeparttable}

\usepackage{multirow}
\usepackage{adjustbox}


%% file: Content/2-Introduction.tex
\section{Introduction}

Motion signals captured from HMDs and handheld controllers contain rich behavioral information that reflects how individuals move and interact in VR and AR applications. Prior work has demonstrated motion-based user identification across diverse XR activities and interaction scenarios, often achieving high performance using head and hand motion trajectories collected from commercial XR devices \cite{miller2020personal, moore2023identifying, nair2023unique, wang2024cross, 10.1145/3411764.3445528, rack2023alyx, liebers2023exploring}. As machine learning becomes increasingly central to XR systems, the demand for large and diverse behavioral datasets continues to grow.

Despite this growing demand, XR behavioral datasets remain relatively small compared with datasets commonly available in other machine learning domains. Most early studies involved fewer than 50 participants \cite{10.1145/3489849.3489880, ajit2019combining, asish2024classification, li2024using, rack2024motion, sivasamy2020vrcauth}, while only a limited number of works have investigated datasets containing more than 100 users \cite{miller2020personal, nair2023inferring, wang2024cross}. In addition, only a subset of these datasets are openly available \cite{moore2024full, rack2023alyx}. Acquiring large-scale XR motion datasets remains costly and time-consuming \cite{fawaz2020deep, tricomi2023you, bulling2014tutorial}, creating a practical barrier to scalable XR behavioral research and motivating alternative approaches for expanding XR behavioral populations.

As a complementary approach to traditional data collection, we explore synthetic behavioral populations by generating synthesized trajectories from existing datasets. We present an interpolation-based motion synthesis pipeline that combines task-aware DTW alignment with trajectory interpolation to generate new trajectories while preserving the temporal structure of the original tasks. Unlike conventional data augmentation, which typically generates additional samples for existing users, our approach generates synthesized trajectories that form synthetic behavioral populations rather than additional samples for existing users.

We apply the proposed pipeline to the publicly available FAST dataset \cite{moore2024full}, generating and openly releasing 100 synthesized trajectories. \footnote{The 100 synthesized XR behavioral trajectories generated in this work are publicly available at:
\url{https://huggingface.co/datasets/XraiSyntheticInterpolations/100_FAST_Synths}.} Motion-based user identification experiments show that hybrid datasets achieve performance comparable to similarly sized real-only datasets while maintaining relatively low confusion between synthesized trajectories and their contributing participants.

Our findings suggest that synthetic behavioral populations can complement traditional data collection and support XR behavioral research at scales beyond those feasible through direct data collection alone.

The contributions of this paper are as follows:
\begin{itemize}
\item An interpolation-based XR motion synthesis pipeline that combines task-aware temporal alignment and motion interpolation to generate synthetic behavioral trajectories from existing XR datasets.
\item An openly available synthetic XR behavioral dataset containing 100 trajectories across two VR assembly tasks, derived from a large-scale public XR dataset \cite{moore2024full}, enabling research on synthetic behavioral populations for immersive systems.
\item An empirical evaluation demonstrating that synthesized trajectories preserve meaningful behavioral structure and can support the creation of larger synthetic behavioral populations.
\end{itemize}

%% file: Content/3-Related_Work.tex
\section{Related Work}
\input{Tables/RW_models}

Prior research on XR motion-based user identification spans diverse activities, datasets, and modeling approaches. Table~\ref{tab:related_work_comparison} summarizes representative studies from the literature. Because the present work focuses on expanding existing XR motion datasets through motion synthesis, we review prior work from two perspectives: the characteristics and availability of XR motion datasets, and the machine learning methods commonly used to evaluate XR motion data.

\subsection{XR Motion Datasets and Identification}

XR motion-based user identification has been investigated across a wide range of immersive activities and interaction scenarios. Existing studies have explored activities including object manipulation \cite{olade2020biomove}, target selection \cite{pfeuffer2019behavioural}, ball throwing \cite{miller2020within, li2024using}, gameplay \cite{rack2023alyx, liebers2023exploring}, and VR assembly tasks \cite{moore2023identifying, wang2024cross}. Most studies rely on motion signals captured from the head and handheld controllers \cite{liebers2023exploring, nair2023unique, nair2023inferring}, while a smaller subset investigates head-only or single-hand interaction settings \cite{mustafa2018unsure, sivasamy2020vrcauth, kupin2018task}.

Dataset availability has strongly influenced which XR activities have been repeatedly studied. As summarized in Table~\ref{tab:related_work_comparison}, several publicly available datasets, particularly those associated with ``Half-Life: Alyx'' \cite{rack2023alyx} and ``Beat Saber'' \cite{boxrr23_dataset}, have been reused across multiple studies. In contrast, many other datasets appear only once in the literature, and some datasets reported as open are no longer publicly accessible.

Collecting large-scale XR motion datasets remains expensive and time-consuming due to participant recruitment, experimental setup requirements, and the effort required to capture, process, and manage large volumes of motion data \cite{fawaz2020deep, bulling2014tutorial}. These challenges motivate the exploration of approaches that can expand existing XR motion datasets without requiring additional large-scale user studies. In this work, we investigate whether interpolation-based motion synthesis can generate additional motion trajectories that support larger--scale machine learning experiments using existing XR datasets.

\subsection{Machine Learning Methods in XR}

The data pre-processing and modeling approaches summarized in Table~\ref{tab:related_work_comparison} reveal several recurring patterns. Among them, ``1s intervals + summary statistics'' is one of the most commonly used approaches, appearing in multiple studies \cite{liebers2023exploring, moore2023identifying, moore2021personal, wang2024cross}. Other frequently used techniques include pairwise trajectory matching \cite{ajit2019combining, kupin2018task, miller2020within}, sequence-length standardization \cite{asish2022user, liebers2024kinetic}, handcrafted feature extraction \cite{sivasamy2020vrcauth}, and rolling or sliding-window segmentation strategies \cite{mustafa2018unsure, rack2023alyx, rack2023versatile}.

A variety of machine learning models have been applied to XR motion data, including Random Forest \cite{liebers2024identifying, liebers2023exploring}, k-nearest neighbors \cite{asish2022user, olade2020biomove}, gradient boosting \cite{moore2023identifying, nair2023unique, wang2024cross}, recurrent neural networks \cite{10.1145/3411764.3445528}, convolutional neural networks \cite{rack2023alyx}, and transformer-based models \cite{li2024using, nair2023inferring}. Within studies using ``1s intervals + summary statistics,'' gradient boosting machines (GBM) have repeatedly achieved strong performance across multiple XR datasets, including assembly tasks \cite{moore2023identifying, wang2024cross}and Beat Saber \cite{liebers2023exploring, nair2023inferring, nair2023unique}. Therefore, the present work adopts the same preprocessing strategy and classifier to facilitate comparison with prior XR identification research.

%% file: Tables/RW_models.tex
\begin{table*}[!t]
\centering
\caption{Comparison of prior work and our study in terms of activity, sensing modality, number of trajectories, dataset availability, data pre-processing, and best-performing model. All best performance values are rounded to two significant digits.}
\label{tab:related_work_comparison}

\scriptsize
\setlength{\tabcolsep}{1.8pt}
\renewcommand{\arraystretch}{1.05}
\resizebox{\textwidth}{!}{%
\begin{tabular}{l l c c c c c c l c c}
\toprule
\multirow{2}{*}{Ref} &
\multirow{2}{*}{Activity} &
\multicolumn{2}{c}{Dataset} &
\multicolumn{4}{c}{Devices} &
\multicolumn{3}{c}{Best Model} \\
\cmidrule(lr){3-4}
\cmidrule(lr){5-8}
\cmidrule(lr){9-11}
& & N & Open & Head & 2 Hands & 1 Hand & Eye & Pre-processing & Algorithm & Best Perf. \\
\midrule
\cite{10.1145/3489849.3489880} 
& Tracking moving stimuli 
& 11 & ?
& \checkmark & & & \checkmark 
& 2.5s rolling windows 
& Encoder 
& 1.00 \\

\cite{olade2020biomove} 
& Moving objects to bin 
& 15 & 
& \checkmark & & \checkmark & \checkmark 
& Linear interpolation + PCA 
& kNN & 0.99 \\

\cite{liebers2023exploring} 
& Playing ``Beat Saber" 
& 15 & \cite{boxrr23_dataset} 
& \checkmark & \checkmark & & 
& Handcrafted features 
& RF & 0.81 \\

\cite{liebers2024identifying} 
& UI interaction 
& 16 &  
& \checkmark & \checkmark & & 
& Summary statistics 
& RF & 0.95 \\

\cite{10.1145/3411764.3445528} 
& Shooting an arrow 
& 16 & ?
& \checkmark & \checkmark & & 
& Rolling Windows 
& RNN & 0.90 \\

\cite{pfeuffer2019behavioural} 
& Target pointing 
& 19 & 
& \checkmark & \checkmark & & \checkmark
& Task segmentation 
& RF & 0.64 \\

\cite{mustafa2018unsure} 
& Path navigation 
& 23 & 
& \checkmark & & & 
& 12s rolling windows + PCA 
& SVMs & 0.93 \\

\cite{10.1145/3334480.3382799} 
& Cube tapping
& 23 & 
& & \checkmark & & 
& Multivariate Time Series 
& FCN & 0.99 \\

\cite{liebers2024kinetic} 
& Playing sports
& 24 &
& \checkmark & \checkmark & & 
& Sequence-length standardization 
& FCN & 0.75 \\

\cite{ajit2019combining} 
& Ball throwing 
& 33 & 
& \checkmark & \checkmark & & 
& Pairwise trajectory matching
& Perceptron & 0.93 \\

\cite{kupin2018task} 
& Ball throwing 
& 33 & 
& & & \checkmark & 
& Pairwise trajectory matching
& kNN & 0.93 \\

\cite{asish2022user} 
& Viewing education scenario 
& 34 & 
& \checkmark & & \checkmark & 
& Sequence-length standardization
& kNN & 1.00 \\

\cite{rack2022comparison} 
& Talking with hands 
& 34 & \cite{talkingwithhands32m} 
& \checkmark & \checkmark & & 
& 3s sliding windows 
& LSTM & 1.00 \\

\cite{tricomi2023you} 
& VR robot teleoperation 
& 35 &
& \checkmark & \checkmark & & \checkmark
& 5-timestep Motion derivatives 
& LR & 0.80 \\

& AR everyday application 
& 35 &
& \checkmark & & &
& 5-timestep Motion derivatives 
& LR & 0.97 \\

\cite{miller2020within} 
& Throwing a ball 
& 41 & ? 
& \checkmark & \checkmark & & 
& Pairwise trajectory matching 
& Perceptron & 0.97\\

\cite{9756791} 
& Throwing a ball 
& 41 & ? 
& \checkmark & \checkmark & & 
& Pairwise trajectory matching 
& SNN & 0.97\\

\cite{li2024using} 
& Throwing a ball
& 41 & ? 
& & & \checkmark & 
& Variable rolling windows 
& Transformer & 0.98\\

\cite{moore2023identifying} 
& Assembling structures 
& 45 & \cite{moore2023identifying}
& \checkmark & \checkmark & &  
& 1s intervals + Summary statistics 
& GBM & 0.96 \\

\cite{sivasamy2020vrcauth} 
& Driving simulator
& 40 & \cite{vr_driving_dataset} 
& \checkmark & & & 
& Handcrafted features 
& LMT & 1.00 \\

& Video streaming 
& 48 & \cite{wu_vr_dataset} 
& \checkmark & & & 
& Handcrafted features 
& PART & 1.00 \\

\cite{rack2024motion} 
& Motion password 
& 48 & \cite{mops_dataset}
& \checkmark & \checkmark & &  
& Sequence modeling 
& GRU+Transformer & 0.93 \\

\cite{moore2021personal} 
& Troubleshooting a robot 
& 60 &
& \checkmark & \checkmark & & 
& 1s intervals + Summary statistics
& RF & 0.96 \\

\cite{rack2023versatile} 
& Playing ``Half-Life: Alyx" 
& 63 & \cite{rack2023alyx} 
& \checkmark & \checkmark & & 
& 33s rolling windows 
& GRU & 0.98 \\

\cite{rack2023alyx} 
& Playing ``Half-Life: Alyx" 
& 71 & \cite{rack2023alyx} 
& \checkmark & \checkmark & & 
& 20s rolling windows 
& CNN & 0.90 \\

\cite{wang2024cross} 
& Playing ``Half-Life: Alyx" 
& 63 & \cite{rack2023alyx}
& \checkmark & \checkmark & &  
& 1s intervals + Summary statistics 
& kNN & 0.89 \\

& Assembling structures 
& 106 & \cite{moore2024full} 
& \checkmark & \checkmark & & 
& 1s intervals + Summary statistics 
& GBM & 0.85 \\

\cite{miller2023large} 
& Talking in groups 
& 232 &  
& \checkmark & \checkmark & & 
& Variable sliding windows
& RF & 0.76 \\

\cite{nair2023inferring} 
& Playing ``Beat Saber" 
& 311 & \cite{boxrr23_dataset} 
& \checkmark & \checkmark & & 
& Sequence-length standardization 
& Transformer & 0.92 \\

\cite{miller2020personal} 
& 360 video viewing 
& 511 & 
& \checkmark & & & 
& 1s intervals + Summary statistics
& RF & 0.95 \\

\cite{nair2023unique} 
& Playing ``Beat Saber" 
& 50k+ & \cite{boxrr23_dataset} 
& \checkmark & \checkmark & & 
& 1s intervals + Summary statistics 
& GBM & 0.97 \\

\midrule

\textbf{Ours} & \textbf{Assembling structures} & \textbf{200} & \textbf{*} & \textbf{\checkmark} & \textbf{\checkmark} & & & \textbf{1s intervals + Summary statistics } & \textbf{GBM} & 0.98 \\

\bottomrule
\end{tabular}%
}

\vspace{2pt}
\raggedright
\footnotesize


\textit{?: } Some datasets are reported as publicly available by the authors but could not be located at the time of writing.\\
\textit{*: } Openly available.\\

\textit{All algorithms: } 
kNN = k-nearest Neighbors,
RF = Random Forest, 
RNN = Recurrent Neural Network,
SVM = Support Vector Machine, 
FCN = Fully Convolutional Network,
LSTM = Long Short-Term Memory network,
LR = Logistic Regression, 
SNN = Siamese Neural Network,
GBM = Gradient Boosting Machine, 
LMT = Logistic Model Tree, 
PART = Partial Decision Tree Rule Learner, 
GRU = Gated Recurrent Unit, 
CNN = Convolutional Neural Network.\\

\textit{Other Abbr.: } 
PCA = Principal Component Analysis. \\

\end{table*}

%% file: Content/4-Identity-Level_Data_Synthesis.tex
\section{XR Motion Synthesis Pipeline}

\begin{figure}[t]
\centering
\Description{Pipeline diagram showing two source trajectories, substep segmentation, substep-wise DTW alignment, motion synthesis, temporal reconstruction, and the final synthesized trajectory.}
\includegraphics[width=1.0\linewidth]{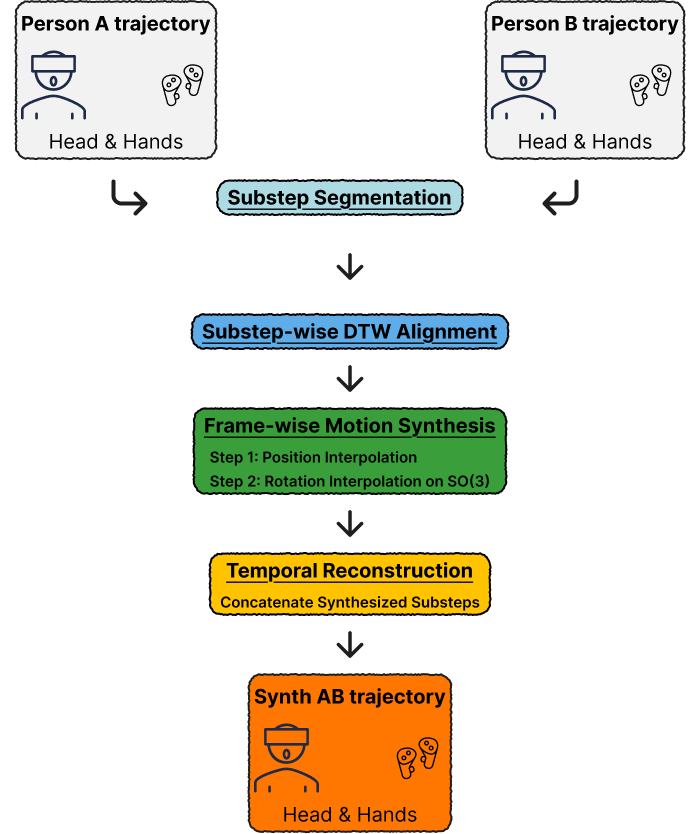}
\caption{Overview of the interaction synthesis pipeline. Two source trajectories, $Person~A$ and $Person~B$, are first segmented into ordered assembly substeps and temporally aligned using DTW. Each aligned substep is then synthesized through frame-wise position interpolation and geodesic rotation interpolation on $SO(3)$, and the synthesized substeps are reconstructed into a continuous trajectory $Syn~AB$.}
\label{fig:Synth_Pipeline}
\end{figure}

Figure~\ref{fig:Synth_Pipeline} provides an overview of the proposed XR motion synthesis pipeline. Starting from pairs of real XR motion trajectories, the pipeline performs substep-level temporal alignment, synthesizes positions and orientations through interpolation, and reconstructs the resulting substeps into continuous motion trajectories.

\subsection{Source Dataset}

We used the FAST dataset \cite{moore2024full}, which contains motion recordings from 108 participants (56 males, 50 females, and 2 non-binary) performing two VR assembly tasks (FAST A and FAST B) using virtual FunPhix construction toys. Head and hand motion data were recorded at 90 Hz using an HTC Vive Pro Eye HMD and handheld controllers. The original FAST dataset publication did not report additional demographic statistics such as participant age.

Because the synthesis pipeline paired within gender groups, the two non-binary participants were not included in the synthesis procedure. Next, motion sequences for each participant and task were segmented into fine-grained assembly substeps (e.g., using a key or attaching components). To ensure reliable DTW alignment during synthesis, we examined the duration of each substep and removed participants whose sequences contained excessively short substeps. After filtering, the final dataset consists of 100 participants (46 females and 54 males).

\subsection{Motion Synthesis}

Additional motion trajectories were generated by pairing participants within gender groups using a cyclic pairing strategy. For each pair, corresponding assembly substeps were temporally aligned using DTW \cite{berndt1994using} applied to right-hand position trajectories. Performing alignment separately within each assembly substep preserved the temporal structure of the assembly task while accommodating differences in execution speed between participants.

\begin{figure}[t]
\centering
\Description{Line plots showing head, left-hand, and right-hand position trajectories for two source participants and one synthesized trajectory. Columns correspond to the $X$, $Y$, and $Z$ spatial axes, and rows correspond to tracked devices. The synthesized trajectory follows patterns observed in the source trajectories.}
\includegraphics[width=1.0\linewidth]{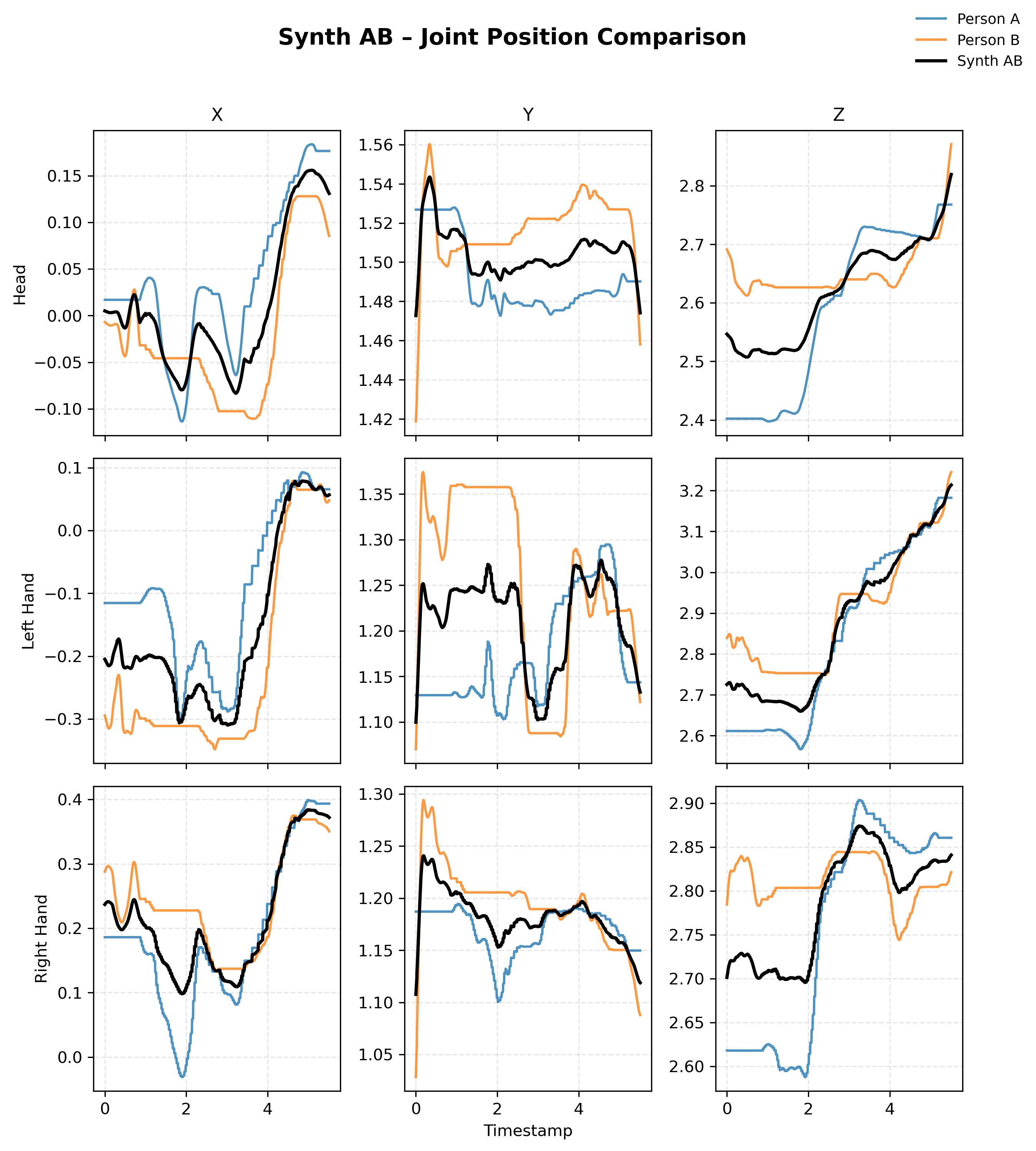}
\caption{Joint position trajectories of the head, left hand, and right hand for two real participants (Person $A$ and Person $B$) and the synthesized motion sequence (Synth $AB$). Columns correspond to the $X$, $Y$, and $Z$ spatial axes, while rows correspond to tracked devices.}
\label{fig:joint_position}
\end{figure}

After alignment, positions from the head and both controllers were synthesized using frame-wise linear interpolation. A fixed interpolation coefficient of $\alpha=0.5$ was used to assign equal contribution to both source trajectories and generate motion samples centered between the two inputs:

\begin{equation}
P_S = (1-\alpha)P_A + \alpha P_B .
\end{equation}

where $P_A$ and $P_B$ denote aligned position samples from the two source trajectories and $P_S$ denotes the synthesized position. Figure~\ref{fig:joint_position} illustrates an example synthesized trajectory. Across all tracked devices, the synthesized motion follows the overall spatial trends of the two source trajectories while remaining smoothly interpolated between them.

Device orientations were synthesized using quaternion log--exp interpolation \cite{grassia1998practical} on $SO(3)$. Using the same interpolation coefficient ($\alpha=0.5$), synthesized orientations were computed as

\begin{equation}
q_S=q_A\exp(\alpha\log(q_A^{-1}q_B)).
\end{equation}

\begin{figure*}[t]
\centering
\Description{Line plots showing right-hand quaternion trajectories over time for two source participants and one synthesized trajectory. The synthesized trajectory follows intermediate rotational patterns between the source trajectories.}
\includegraphics[width=0.75\linewidth]{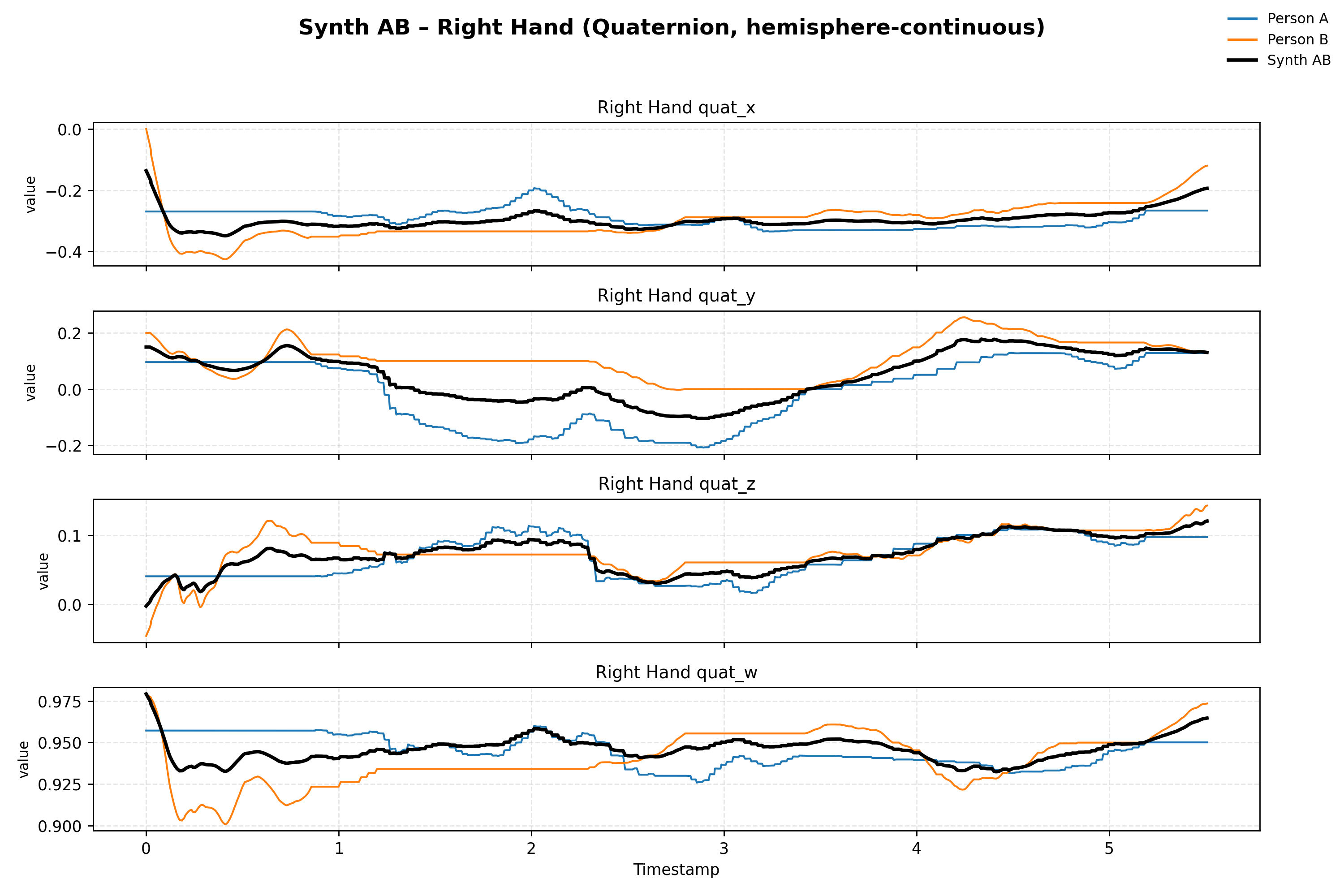}
\caption{Right-hand quaternion trajectories for two real participants and the synthesized motion.}
\label{fig:joint_quaternions}
\end{figure*}

The synthesized orientations were subsequently converted to the continuous 6D rotation representation \cite{zhou2019continuity} used in the machine learning analysis. Figure~\ref{fig:joint_quaternions} shows an example synthesized orientation trajectory for the right-hand controller.

\subsection{Trajectory Reconstruction}

The synthesized assembly substeps were reconstructed into complete task trajectories according to their original assembly order. For each synthesized substep, the target duration was defined as the average duration of the corresponding source substeps, ensuring equal temporal contribution from both source trajectories. The reconstructed trajectory therefore preserved both the ordering of the assembly task and the interpolated motion characteristics derived from the aligned source trajectories.

%% file: Content/5-Machine_Learning_Experiment.tex
\section{Machine Learning Evaluation}

\subsection{Feature Representation}

We evaluated whether the synthesized trajectories retained distinguishable motion characteristics using a user--identification task. Motion features were extracted from the HMD and two handheld controllers in the FAST dataset. For each device, we used 3D position and the continuous 6D rotation representation \cite{zhou2019continuity}, resulting in 27 motion features per frame.

Following prior XR identification work \cite{miller2020personal, moore2021personal, moore2023identifying}, motion trajectories were segmented into one-second intervals and summarized using five statistical descriptors (maximum, minimum, median, mean, and standard deviation) for each feature. The resulting feature vectors were used as input to a GBM classifier \cite{nair2023unique, asish2022user, wang2024cross}.

\subsection{Experimental Design}

To compare hybrid and real--only trajectory datasets, we constructed three real--only datasets containing 24, 48, and 100 trajectories and four hybrid datasets containing 24, 48, 100, and 200 total trajectories. Hybrid datasets combined real participants with synthesized trajectories generated by the proposed pipeline.

For each hybrid dataset, synthesized trajectories were selected first. The corresponding source participants used to generate these motions were then retrieved from the synthesis records, and duplicate real trajectories were removed to obtain the final set of real users.

Identification experiments were conducted separately on the FAST A and FAST B tasks. For each participant, motion data were divided into ten consecutive sub-sessions. We adopted a nested Monte Carlo protocol with 20 runs. In each run, one sub-session was used for testing, one for validation, and the remaining eight for training.

GBM hyperparameters were optimized using Optuna with 40 trials. The search space included learning rate, tree depth, child weight, feature and instance subsampling ratios, L1 and L2 regularization parameters, and the number of estimators. Hyperparameter optimization used only the training and validation data, while the test data remained completely unseen during model selection.

We report accuracy, precision, recall, and macro-F1. Reported results correspond to the average test performance across all runs.

%% file: Content/6-Study.tex
\section{Results}
We first compare identification performance between hybrid and real--only datasets, followed by an analysis of misidentification patterns involving synthesized trajectories and their contributing participants.

\subsection{Identification Performance}

\input{Tables/Results_coreMetrics}

Table~\ref{tab:real_vs_hybrid} summarizes identification performance across all dataset configurations. Across both assembly tasks, hybrid datasets achieved performance comparable to similarly sized real--only datasets. For example, in the 100--trajectory dataset, the hybrid dataset achieved accuracies of 92\% and 91\% for FAST A and FAST B, respectively, compared to 93\% and 94\% for the corresponding real--only dataset. Similar trends were observed for the 24-- and 48--trajectory datasets, suggesting that synthesized trajectories can be incorporated into larger datasets without substantially affecting identification performance.

\input{Tables/Results_Errors}

\subsection{Misidentification Analysis}

Table~\ref{tab:error_analysis} summarizes misidentification patterns within the hybrid datasets. We distinguish between source-related confusions, where synthesized motions are confused with the real participants from which they were generated, and errors involving unrelated other motion trajectories.

Across all experimental conditions, confusion between synthesized motions and their source participants remained relatively low. Real-to-synth misidentification rates ranged from 1.0\% to 3.5\%, while synth-to-real misidentification rates ranged from 1.5\% to 6.9\%. Errors involving unrelated trajectories were typically below 1.5\%.

These results indicate that the synthesized trajectories retain motion characteristics similar to their source participants while remaining sufficiently distinguishable to be recognized as separate trajectories by the classifier. 

%% file: Tables/Results_coreMetrics.tex
\begin{table}[t]
\centering
\scriptsize
\setlength{\tabcolsep}{3pt}

\caption{Comparison of identification performance between real-only and hybrid datasets under different numbers of trajectories. All values are rounded to two significant digits.}
\label{tab:real_vs_hybrid}

\resizebox{\linewidth}{!}{%
\begin{tabular}{r|cccc|cccc}
\hline
& \multicolumn{4}{c|}{\textbf{FAST A}}
& \multicolumn{4}{c}{\textbf{FAST B}} \\
\cline{2-9}
\textbf{Dataset}
& \textbf{Acc} & \textbf{F1} & \textbf{Prec} & \textbf{Rec}
& \textbf{Acc} & \textbf{F1} & \textbf{Prec} & \textbf{Rec} \\
\hline

\textit{Real24}
& 0.98 & 0.97 & 0.97 & 0.98
& 0.98 & 0.98 & 0.98 & 0.98 \\

\textit{Hybrid24}
& 0.91 & 0.89 & 0.88 & 0.91
& 0.88 & 0.84 & 0.82 & 0.88 \\

\hline

\textit{Real48}
& 0.95 & 0.93 & 0.92 & 0.95
& 0.95 & 0.94 & 0.93 & 0.95 \\

\textit{Hybrid48}
& 0.94 & 0.92 & 0.91 & 0.94
& 0.92 & 0.90 & 0.89 & 0.92 \\

\hline
\textit{Real100}
& 0.93 & 0.91 & 0.90 & 0.93 
& 0.94 & 0.92 & 0.91 & 0.94 \\
\textit{Hybrid100}
& 0.92 & 0.90 & 0.89 & 0.92
& 0.91 & 0.88 & 0.87 & 0.91 \\
\hline
\textit{Hybrid200}
& 0.91 & 0.88 & 0.87 & 0.91 
& 0.89 & 0.86 & 0.85 & 0.89 \\
\hline

\end{tabular}%
}
\end{table}

%% file: Tables/Results_Errors.tex
\begin{table*}[t]
\centering
\small
\setlength{\tabcolsep}{4pt}

\caption{Breakdown of identification errors for hybrid datasets under different numbers of trajectories in the FAST A and FAST B assembly tasks. Error types include real users identified as derived synthetic motions ($Real \rightarrow Synth$), synthetic trajectories identified as their contributing real users ($Synth \rightarrow Real$), real users identified as unrelated synthetic trajectories ($Real \rightarrow Other~Synth$), and synthetic trajectories identified as unrelated real users ($Synth \rightarrow Other~Real$). All values are rounded to three significant digits.}
\label{tab:error_analysis}

\resizebox{0.90\linewidth}{!}{%
\begin{tabular}{r|cccc|cccc}
\hline

& \multicolumn{4}{c|}{\textbf{FAST A Errors}}
& \multicolumn{4}{c}{\textbf{FAST B Errors}} \\
\cline{2-9}
& 
& 
& \textbf{Real$\rightarrow$}
& \textbf{Synth$\rightarrow$}
& 
& 
& \textbf{Real$\rightarrow$}
& \textbf{Synth$\rightarrow$} \\
\textbf{Dataset}
& \textbf{Real$\rightarrow$Synth}
& \textbf{Synth$\rightarrow$Real}
& \textbf{Other Synth}
& \textbf{Other Real}
& \textbf{Real$\rightarrow$Synth}
& \textbf{Synth$\rightarrow$Real}
& \textbf{Other Synth}
& \textbf{Other Real} \\
\hline


\textit{Hybrid24}
& 0.035 & 0.019 & 0.004 & 0.000
& 0.017 & 0.069 & 0.004 & 0.002 \\

\textit{Hybrid48}
& 0.022 & 0.014 & 0.005 & 0.002
& 0.016 & 0.032 & 0.003 & 0.002 \\

\textit{Hybrid100}
& 0.019 & 0.017 & 0.002 & 0.004
& 0.012 & 0.028 & 0.010 & 0.006 \\

\textit{Hybrid200}
& 0.010 & 0.014 & 0.008 & 0.010
& 0.011 & 0.015 & 0.015 & 0.012 \\





\hline
\end{tabular}%
}

\end{table*}

%% file: Content/7-Discussion.tex
\section{Discussion}

\subsection{Synthesized Trajectories Retain Discriminative Motion Structure}

The hybrid datasets consistently achieved identification performance comparable to similarly sized real-only datasets across both assembly tasks. Although performance decreased modestly as additional synthesized trajectories were introduced, this behavior is expected because the classification problem becomes substantially more difficult as the number of behavioral trajectories increases. Notably, identification accuracy remained close to 90\% even after doubling the number of trajectories from 100 to 200.

Rather than viewing the slight decrease in accuracy as a limitation of the synthesis method, we interpret it as evidence that the synthesized trajectories introduce meaningful behavioral diversity while preserving the discriminative motion characteristics necessary for user identification. If the synthesized trajectories merely duplicated existing participants, identification performance would likely remain artificially high because the classifier would encounter nearly identical behavioral patterns. Conversely, if the synthesized trajectories failed to preserve meaningful behavioral structure, identification performance would degrade substantially as the synthesized users became indistinguishable from one another. The observed results lie between these two extremes, suggesting that the synthesized trajectories behave as plausible new behavioral identities rather than simple copies or random perturbations.

Our findings indicate that synthetic behavioral populations generated through interpolation-based motion synthesis can complement traditional XR data collection by expanding the diversity of available behavioral populations while maintaining sufficient structure for downstream machine learning tasks.

\subsection{Beyond Data Augmentation}

Although the proposed method shares some similarities with conventional data augmentation in DTW, its objective is fundamentally different. Traditional augmentation techniques generate additional observations belonging to existing classes, such as by perturbing trajectories through noise injection, temporal scaling, or spatial transformations \cite{iwana2021empirical}. These approaches primarily improve classifier robustness without increasing the number of behavioral trajectories represented within a dataset.

In contrast, our pipeline generates new synthetic trajectories by combining motion information from pairs of real participants. The synthesized trajectories are evaluated as separate classes rather than treated as additional observations of existing users. The distinction is particularly relevant for XR behavioral research because it enables researchers to construct larger evaluation datasets and investigate how machine learning models behave as the number of distinguishable behavioral trajectories increases.

We therefore view synthetic behavioral populations not as a replacement for conventional data augmentation, but as a complementary approach for enabling research at larger behavioral population scales. Such an approach could support systematic evaluation of behavioral models and data-driven XR systems at scales that are difficult to obtain through traditional user studies.

%% file: Content/8-Limitation_and_Future_Work.tex
\section{Future Work \& Impact}

Synthetic behavioral populations open several promising future research directions that extend beyond the interpolation pipeline presented in the current work.

In the near term, future research should investigate synthesis strategies that incorporate motion information from more than two participants and allow interpolation weights to vary throughout an interaction. Evaluating synthetic trajectories across a wider range of XR activities, sensing modalities, and downstream learning tasks will help establish when synthetic behavioral populations can reliably complement traditional data collection.

Beyond methodological improvements, synthetic behavioral populations have the potential to reshape how XR behavioral datasets are created and shared. Publicly available synthetic datasets and reproducible synthesis pipelines could enable benchmarking across research groups, support controlled evaluation across different dataset scales and compositions, and reduce barriers for researchers who lack access to large participant studies.

Over the longer term, advances in methods for generating synthetic behavioral populations could change the role of existing XR datasets in behavioral research. Rather than serving only as fixed outcomes of individual data collection efforts, public datasets could provide foundations for constructing new experimental datasets tailored to different research questions and evaluation scales. Researchers could use such datasets to explore questions that are difficult to study through direct data collection alone and identify findings that warrant further validation with real participants.

Our contribution represents an initial step toward a broader research agenda in which synthetic behavioral populations become a reusable research capability for XR. By complementing rather than replacing human participant studies, synthetic behavioral populations could expand the range and scale of questions that XR researchers can systematically investigate.

%% file: Content/9-Conclusion.tex
\section{Conclusion}

Our paper evaluated whether interpolation-based motion synthesis can support machine learning identification in XR. Hybrid datasets containing both real and synthesized trajectories achieved identification performance comparable to similarly sized real-only datasets while showing limited source attribution to the original participants. These results suggest that interpolation-based synthesis can expand XR behavioral datasets without substantially degrading their utility for identification tasks.

More broadly, the generation of synthetic behavioral populations offers a practical direction for scaling XR behavioral datasets when collecting additional participant data is expensive or impractical. Future work will evaluate more synthesis methods, behavioral tasks, and downstream learning applications.

%% file: ref.bib
@article{iwana2021empirical,
  title={An empirical survey of data augmentation for time series classification with neural networks},
  author={Iwana, Brian Kenji and Uchida, Seiichi},
  journal={Plos one},
  volume={16},
  number={7},
  pages={e0254841},
  year={2021},
  publisher={Public Library of Science San Francisco, CA USA}
}

@article{bulling2014tutorial,
  title={A tutorial on human activity recognition using body-worn inertial sensors},
  author={Bulling, Andreas and Blanke, Ulf and Schiele, Bernt},
  journal={ACM Computing Surveys (CSUR)},
  volume={46},
  number={3},
  pages={1--33},
  year={2014},
  publisher={ACM New York, NY, USA}
}

@misc{mops_dataset,
  author       = {Schell, Christian},
  title        = {MoPs: Motion Password Dataset},
  year         = {2024},
  howpublished = {\url{https://github.com/cschell/MoPs}},
  note         = {Accessed: 2026}
}

@misc{wu_vr_dataset,
  author       = {Wu, Chulei},
  title        = {VR Behavioral Dataset and Project Page},
  year         = {2022},
  howpublished = {\url{https://wuchlei-thu.github.io/}},
  note         = {Accessed: 2026}
}

@misc{vr_driving_dataset,
  author       = {Sasan Jafarzadeh},
  title        = {Virtual Reality Driving Dataset},
  year         = {2020},
  howpublished = {\url{https://www.kaggle.com/sasanj/virtual-reality-driving}},
  note         = {Accessed: 2026}
}

@misc{talkingwithhands32m,
  author       = {{Facebook AI Research}},
  title        = {TalkingWithHands32M Dataset},
  year         = {2019},
  howpublished = {\url{https://github.com/facebookresearch/TalkingWithHands32M}},
  note         = {Accessed: 2026}
}

@misc{boxrr23_dataset,
  author       = {{Berkeley Reality Lab}},
  title        = {BOXRR-23: Berkeley Open XR Research Repository},
  year         = {2023},
  howpublished = {\url{https://rdi.berkeley.edu/metaverse/boxrr-23/}},
  note         = {Accessed: 2026}
}

@inproceedings{kupin2018task,
  title={Task-driven biometric authentication of users in virtual reality (VR) environments},
  author={Kupin, Alexander and Moeller, Benjamin and Jiang, Yijun and Banerjee, Natasha Kholgade and Banerjee, Sean},
  booktitle={International conference on multimedia modeling},
  pages={55--67},
  year={2018},
  organization={Springer}
}

@inproceedings{zhou2019continuity,
  title={On the continuity of rotation representations in neural networks},
  author={Zhou, Yi and Barnes, Connelly and Lu, Jingwan and Yang, Jimei and Li, Hao},
  booktitle={Proceedings of the IEEE/CVF conference on computer vision and pattern recognition},
  pages={5745--5753},
  year={2019}
}

@article{grassia1998practical,
  title={Practical parameterization of rotations using the exponential map},
  author={Grassia, F Sebastian},
  journal={Journal of graphics tools},
  volume={3},
  number={3},
  pages={29--48},
  year={1998},
  publisher={Taylor \& Francis}
}

@inproceedings{berndt1994using,
  title={Using dynamic time warping to find patterns in time series},
  author={Berndt, Donald J and Clifford, James},
  booktitle={Proceedings of the 3rd international conference on knowledge discovery and data mining},
  pages={359--370},
  year={1994}
}

@inproceedings{sivasamy2020vrcauth,
  title={Vrcauth: continuous authentication of users in virtual reality environment using head-movement},
  author={Sivasamy, Manimaran and Sastry, VN and Gopalan, NP},
  booktitle={2020 5th international conference on communication and electronics systems (icces)},
  pages={518--523},
  year={2020},
  organization={IEEE}
}

@inproceedings{liebers2024kinetic,
  title={Kinetic signatures: A systematic investigation of movement-based user identification in virtual reality},
  author={Liebers, Jonathan and Laskowski, Patrick and Rademaker, Florian and Sabel, Leon and Hoppen, Jordan and Gruenefeld, Uwe and Schneegass, Stefan},
  booktitle={Proceedings of the 2024 CHI Conference on Human Factors in Computing Systems},
  pages={1--19},
  year={2024}
}

@article{miller2023large,
  title={A large-scale study of personal identifiability of virtual reality motion over time},
  author={Miller, Mark Roman and Han, Eugy and DeVeaux, Cyan and Jones, Eliot and Chen, Ryan and Bailenson, Jeremy N},
  journal={arXiv preprint arXiv:2303.01430},
  year={2023}
}

@article{rack2023versatile,
  title={Versatile user identification in extended reality using pretrained similarity-learning},
  author={Rack, Christian and Kobs, Konstantin and Fernando, Tamara and Hotho, Andreas and Latoschik, Marc Erich},
  journal={arXiv preprint arXiv:2302.07517},
  year={2023}
}

@inproceedings{liebers2023exploring,
  title={Exploring the stability of behavioral biometrics in virtual reality in a remote field study: Towards implicit and continuous user identification through body movements},
  author={Liebers, Jonathan and Burschik, Christian and Gruenefeld, Uwe and Schneegass, Stefan},
  booktitle={Proceedings of the 29th ACM Symposium on Virtual Reality Software and Technology},
  pages={1--12},
  year={2023}
}

@article{liebers2024identifying,
  title={Identifying users by their hand tracking data in augmented and virtual reality},
  author={Liebers, Jonathan and Brockel, Sascha and Gruenefeld, Uwe and Schneegass, Stefan},
  journal={International Journal of Human--Computer Interaction},
  volume={40},
  number={2},
  pages={409--424},
  year={2024},
  publisher={Taylor \& Francis}
}

@inproceedings{ajit2019combining,
  title={Combining pairwise feature matches from device trajectories for biometric authentication in virtual reality environments},
  author={Ajit, Ashwin and Banerjee, Natasha Kholgade and Banerjee, Sean},
  booktitle={2019 IEEE International Conference on Artificial Intelligence and Virtual Reality (AIVR)},
  pages={9--97},
  year={2019},
  organization={IEEE Computer Society}
}

@inproceedings{miller2020within,
  title={Within-system and cross-system behavior-based biometric authentication in virtual reality},
  author={Miller, Robert and Banerjee, Natasha Kholgade and Banerjee, Sean},
  booktitle={2020 IEEE conference on virtual reality and 3D user interfaces abstracts and workshops (VRW)},
  pages={311--316},
  year={2020},
  organization={IEEE}
}

@article{olade2020biomove,
  title={Biomove: Biometric user identification from human kinesiological movements for virtual reality systems},
  author={Olade, Ilesanmi and Fleming, Charles and Liang, Hai-Ning},
  journal={Sensors},
  volume={20},
  number={10},
  pages={2944},
  year={2020},
  publisher={MDPI}
}

@article{tricomi2023you,
  title={You can’t hide behind your headset: User profiling in augmented and virtual reality},
  author={Tricomi, Pier Paolo and Nenna, Federica and Pajola, Luca and Conti, Mauro and Gamberini, Luciano},
  journal={IEEE Access},
  volume={11},
  pages={9859--9875},
  year={2023},
  publisher={IEEE}
}

@inproceedings{nair2023unique,
  title={Unique identification of 50,000+ virtual reality users from head \& hand motion data},
  author={Nair, Vivek and Guo, Wenbo and Mattern, Justus and Wang, Rui and O'Brien, James F and Rosenberg, Louis and Song, Dawn},
  booktitle={32nd USENIX Security Symposium (USENIX Security 23)},
  pages={895--910},
  year={2023}
}

@inproceedings{moore2023identifying,
  title={Identifying virtual reality users across domain-specific tasks: A systematic investigation of tracked features for assembly},
  author={Moore, Alec G and Do, Tiffany D and Ruozzi, Nicholas and McMahan, Ryan P},
  booktitle={2023 IEEE International Symposium on Mixed and Augmented Reality (ISMAR)},
  pages={396--404},
  year={2023},
  organization={IEEE}
}

@inproceedings{moore2021personal,
  title={Personal identifiability and obfuscation of user tracking data from VR training sessions},
  author={Moore, Alec G and McMahan, Ryan P and Dong, Hailiang and Ruozzi, Nicholas},
  booktitle={2021 IEEE International Symposium on Mixed and Augmented Reality (ISMAR)},
  pages={221--228},
  year={2021},
  organization={IEEE}
}

@article{miller2020personal,
  title={Personal identifiability of user tracking data during observation of 360-degree VR video},
  author={Miller, Mark Roman and Herrera, Fernanda and Jun, Hanseul and Landay, James A and Bailenson, Jeremy N},
  journal={Scientific Reports},
  volume={10},
  number={1},
  pages={17404},
  year={2020},
  publisher={Nature Publishing Group UK London}
}

@inproceedings{pfeuffer2019behavioural,
  title={Behavioural biometrics in vr: Identifying people from body motion and relations in virtual reality},
  author={Pfeuffer, Ken and Geiger, Matthias J and Prange, Sarah and Mecke, Lukas and Buschek, Daniel and Alt, Florian},
  booktitle={Proceedings of the 2019 CHI Conference on Human Factors in Computing Systems},
  pages={1--12},
  year={2019}
}

@inproceedings{wang2024cross,
  title={Cross-Domain Gender Identification Using VR Tracking Data},
  author={Wang, Qidi J and Moore, Alec G and Chawla, Nayan N and McMahan, Ryan P},
  booktitle={2024 IEEE International Symposium on Mixed and Augmented Reality (ISMAR)},
  pages={180--189},
  year={2024},
  organization={IEEE}
}

@article{fawaz2020deep,
  title={Deep learning for time series classification},
  author={Fawaz, Hassan Ismail},
  journal={arXiv preprint arXiv:2010.00567},
  year={2020}
}

@article{moore2024full,
  title={The Full-scale Assembly Simulation Testbed (FAST) Dataset},
  author={Moore, Alec G and Do, Tiffany D and Chawla, Nayan N and Iriarte, Antonia Jimenez and McMahan, Ryan P},
  journal={arXiv preprint arXiv:2403.08969},
  year={2024}
}

@article{rack2023alyx,
  title={Who is Alyx? A new behavioral biometric dataset for user identification in XR},
  author={Rack, Christian and Fernando, Tamara and Yalcin, Murat and Hotho, Andreas and Latoschik, Marc Erich},
  journal={Frontiers in Virtual Reality},
  volume={4},
  pages={1272234},
  year={2023},
  publisher={Frontiers Media SA}
}

@inproceedings{li2024using,
  title={Using Motion Forecasting for Behavior-Based Virtual Reality (VR) Authentication},
  author={Li, Mingjun and Banerjee, Natasha Kholgade and Banerjee, Sean},
  booktitle={2024 IEEE International Conference on Artificial Intelligence and eXtended and Virtual Reality (AIxVR)},
  pages={31--40},
  year={2024},
  organization={IEEE}
}

@inproceedings{rack2024motion,
  title={Motion Passwords},
  author={Rack, Christian and Schach, Lukas and Achter, Felix and Shehada, Yousof and Lin, Jinghuai and Latoschik, Marc Erich},
  booktitle={Proceedings of the 30th ACM Symposium on Virtual Reality Software and Technology},
  pages={1--11},
  year={2024}
}

@article{asish2024classification,
  title={Classification of Internal and External Distractions in an Educational VR Environment Using Multimodal Features},
  author={Asish, Sarker M and Kulshreshth, Arun K and Borst, Christoph W and Sutradhar, Shaon},
  journal={IEEE Transactions on Visualization and Computer Graphics},
  year={2024},
  publisher={IEEE}
}

@inproceedings{asish2022user,
  title={User identification utilizing minimal eye-gaze features in virtual reality applications},
  author={Asish, Sarker Monojit and Kulshreshth, Arun K and Borst, Christoph W},
  booktitle={Virtual Worlds},
  volume={1},
  number={1},
  pages={42--61},
  year={2022},
  organization={MDPI}
}

@inproceedings{mustafa2018unsure,
  title={Unsure how to authenticate on your vr headset? come on, use your head!},
  author={Mustafa, Tahrima and Matovu, Richard and Serwadda, Abdul and Muirhead, Nicholas},
  booktitle={Proceedings of the Fourth ACM International Workshop on Security and Privacy Analytics},
  pages={23--30},
  year={2018}
}

@inproceedings{10.1145/3334480.3382799,
author = {Mathis, Florian and Fawaz, Hassan Ismail and Khamis, Mohamed},
title = {Knowledge-driven Biometric Authentication in Virtual Reality},
year = {2020},
isbn = {9781450368193},
publisher = {Association for Computing Machinery},
address = {New York, NY, USA},
url = {https://doi-org.ezproxy.lib.vt.edu/10.1145/3334480.3382799},
doi = {10.1145/3334480.3382799},
booktitle = {Extended Abstracts of the 2020 CHI Conference on Human Factors in Computing Systems},
pages = {1–10},
numpages = {10},
location = {Honolulu, HI, USA},
series = {CHI EA '20}
}

@inproceedings{10.1145/3411764.3445528,
author = {Liebers, Jonathan and Abdelaziz, Mark and Mecke, Lukas and Saad, Alia and Auda, Jonas and Gruenefeld, Uwe and Alt, Florian and Schneegass, Stefan},
title = {Understanding User Identification in Virtual Reality Through Behavioral Biometrics and the Effect of Body Normalization},
year = {2021},
isbn = {9781450380966},
publisher = {Association for Computing Machinery},
address = {New York, NY, USA},
url = {https://doi-org.ezproxy.lib.vt.edu/10.1145/3411764.3445528},
doi = {10.1145/3411764.3445528},
booktitle = {Proceedings of the 2021 CHI Conference on Human Factors in Computing Systems},
articleno = {517},
numpages = {11},
location = {Yokohama, Japan},
series = {CHI '21}
}

@INPROCEEDINGS{9756791,
  author={Miller, Robert and Banerjee, Natasha Kholgade and Banerjee, Sean},
  booktitle={2022 IEEE Conference on Virtual Reality and 3D User Interfaces (VR)}, 
  title={Combining Real-World Constraints on User Behavior with Deep Neural Networks for Virtual Reality (VR) Biometrics}, 
  year={2022},
  volume={},
  number={},
  pages={409-418},
  doi={10.1109/VR51125.2022.00060}}

@inproceedings{rack2022comparison,
  title={Comparison of data encodings and machine learning architectures for user identification on arbitrary motion sequences},
  author={Rack, Christian and Hotho, Andreas and Latoschik, Marc Erich},
  booktitle={2022 IEEE International Conference on Artificial Intelligence and Virtual Reality (AIVR)},
  pages={11--19},
  year={2022},
  organization={IEEE}
}

@article{nair2023inferring,
  title={Inferring private personal attributes of virtual reality users from head and hand motion data},
  author={Nair, Vivek and Rack, Christian and Guo, Wenbo and Wang, Rui and Li, Shuixian and Huang, Brandon and Cull, Atticus and O'Brien, James F and Latoschik, Marc and Rosenberg, Louis and others},
  journal={arXiv preprint arXiv:2305.19198},
  year={2023}
}

@inproceedings{10.1145/3489849.3489880,
author = {Liebers, Jonathan and Horn, Patrick and Burschik, Christian and Gruenefeld, Uwe and Schneegass, Stefan},
title = {Using Gaze Behavior and Head Orientation for Implicit Identification in Virtual Reality},
year = {2021},
isbn = {9781450390927},
publisher = {Association for Computing Machinery},
address = {New York, NY, USA},
url = {https://doi-org.ezproxy.lib.vt.edu/10.1145/3489849.3489880},
doi = {10.1145/3489849.3489880},
booktitle = {Proceedings of the 27th ACM Symposium on Virtual Reality Software and Technology},
articleno = {22},
numpages = {9},
location = {Osaka, Japan},
series = {VRST '21}
}
